\documentclass[12pt]{amsart}
\usepackage{geometry} % see geometry.pdf on how to lay out the page. There's lots.
\usepackage[utf8]{inputenc}
\usepackage{amsmath}
\usepackage{amsfonts}
\usepackage{amssymb}
\usepackage{graphicx}
\usepackage{tabularx}
\usepackage{booktabs}
\title{Mathematical modelling of fractional order circuits}
\author{Miguel Angel Moreles \and Rafael Lainez}
\address{CIMAT, Jalisco S/N, Valenciana, 
Guanajuato GTO M\'exico 36023}
\email{moreles@cimat.mx, rafael.lainez@cimat.mx}

\date{} % delete this line to display the current date

\begin{document}

\maketitle

\begin{abstract}
In this work a classical derivation of fractional order circuits models is presented. Generalized constitutive equations in terms of fractional Riemann-Liouville derivatives are introduced in the Maxwell's equations. Next the Kirchhoff voltage law is applied in a RCL circuit configuration.  A fractional differential equation model is obtained with Caputo derivatives. Thus standard initial conditions apply.
\end{abstract}

\tableofcontents

\section{Introduction}

In the las decades there has been a great interest on Fractional Calculus and its applications. Models involving fractional derivatives and operators have been found to better describe some real phenomena. A historical review of applications is presented in Machado et al \cite{Machadoetal}.

In this work we are interested on applications to linear systems, in particular to fractional circuits models. Research on the topic is very active and lively.
Form the perspective of Systems Theory, an inviting introduction to Fractional Systems is presented in Duarte \cite{Duarte}. See also the work of Radwan \& Salama \cite{RadwanSalama}. Focusing on the applications to fractional order circuits is the work of Elwakil \cite{Elwakil}. Therein the motivation  for fractional order circuits from Biochemistry and Medicine is described.
A more extensive survey on models from Biology and Biomedicine for fitting impedance data is presented in Freeborn \cite{Freeborn}. With regards to the 
Cole Model, the Laplace transform model of a fractional order circuit, it is pointed out that : \emph{"while this model is effective at representing experimentally collected bioimpedance data, it does not provide an explanation of the underlying mechanisms".}

Our modest aim is to provide some insight on this issue. Our purpose is to show that the fractional order circuit model can be obtained from a classical modelling approach. We consider the RCL circuit configuration  and apply the Maxwell's equations as customary. The novelty is on introducing generalised constitutive equations relating the electric flux density, the magnetic field intensity, and the electric current density with the electric and magnetic fields.
These laws are in terms of fractional time derivatives in the sense of Riemann-Liouville. In particular, a generalised Ohm's law is introduced which is consistent with the empirical Curie's law. This law is the point of departure of our work, it is fully discussed in Westerlund \& Ekstam \cite{Cour}. 

The fractional order circuit model to be obtained it is, as expected, a fractional ordinary differential equation in the Riemann-Liouville sense. The underlying mechanisms of physical phenomena where this model applies, are partially explained by the generalised constitutive laws to be introduced.

It is well known that initial conditions need not be well defined in the Riemann-Liouville case. We show that It is straightforward to obtain an equivalent model  with derivatives in the Caputo sense. Consequently, initial conditions can be prescribed as usual. 

The outline is as follows.

In Section 2 we list some basic definitions and properties of Fractional Calculus, these shall be used freely throughout. In Section 3 we review Curie's law and recall the Maxwell's equations in continuous media. From Curie's law we make a case for a generalised Ohm's law in lossy media. The classical modelling of RCL circuits is carried out in Section 4. Then, generalized laws for the electric flux density and magnetic field intensity are introduced leading to fractional versions of circuit componentes. In Section 5 we apply the Kirchhofff's law of voltages to obtain the desired fractional order circuit model.

\section{Preliminaries on Fractional Calculus}

For later reference let us recall some basic properties of Fractional Calculus, see Diethelm \cite{Frac}. 

Let $s>0$ and $f:\mathbb{R}\rightarrow\mathbb{R}$.  The Riemann-Liouville fractional integral of order $s$ centered at $0$, $J^s f$ is given by:

\begin{align}
J^s f(t)=\frac{1}{\Gamma(s)}\int_0^t\left(t-\sigma\right)^{s-1}f(\sigma)d\sigma.
\end{align}

Hereafter, let $0<\alpha\leq 1$. The Riemann-Liouville derivative $D^\alpha f(x)$ of order $\alpha$ centered at $0$ is defined by

\begin{align}
D^\alpha f=DJ^{1-\alpha}f.
\end{align}

An alternative definition is given by Caputo, $\hat{D}^\alpha f$ , namely
\begin{align*}
\hat{D}^{\alpha}f=J^{1-\alpha}Df.
\end{align*}
\bigskip

Let $\mathbb{R}^+=\lbrace t\in \mathbb{R}: t>0\rbrace$ and $\mathbb{R}^+_0=\lbrace t\in \mathbb{R}: t\geq 0\rbrace$. Let us denote by $C^0(\mathbb{R}^+)$  and $C^0(\mathbb{R}^+_0)$ spaces of continuous functions Also let $L^1_{loc}(\mathbb{R}^+)$ be the space of locally integrable functions. 

\bigskip

The following are well known facts about derivatives and integrals  in the sense of Riemann-Liouville:

\begin{itemize}
\item $J^rJ^s=J^{r+s},\quad r,s>0.\quad$ (semigroup property)

\smallskip

\item For $\lambda>-1$
\[
D^\alpha  t^\lambda=\frac{\Gamma(\lambda+1)}{\Gamma(\lambda-\alpha+1)}t^{\lambda-\alpha},
\]
giving in particular $D^\alpha t^{\alpha-1}=0$.

\smallskip

\item $D^{\alpha}J^{\alpha} f = f\quad$ for all $f\in C^0(\mathbb{R}^+)\cap L^1_{loc}(\mathbb{R}^+).$ 

\smallskip

\item If $u\in C^0(\mathbb{R}^+)\cap L^1_{loc}(\mathbb{R}^+),$ then the fractional differential equation
\[
D^\alpha u=0
\]
has $u=ct^{\alpha-1},$ $c\in\mathbb{R}$, as unique solutions.

\smallskip

\item If $f, D^{\alpha}f\in C^0(\mathbb{R}^+)\cap L^1_{loc}(\mathbb{R}^+)$ then 
\[
J^{\alpha}D^{\alpha} f(t)=f(t)+ct^{\alpha-1}
\]
for some $c\in\mathbb{R}$. When $f\in C^0(\mathbb{R}^+_0),$ $c=0$.

\end{itemize}

\bigskip

Our aim is to stress the ideas for the derivation of a fractional circuit model. Thus, we shall assume that the functions under consideration are sufficiently regular. It will become apparent that, with some technicalities, the arguments that follow are valid for functions with weaker properties.

\bigskip

For instance, consider the space $C_r^0(\mathbb{R}^+),$ $r\geq 0$ of functions $f\in C^0(\mathbb{R}^+)$ such that $t^r f\in C^0(\mathbb{R}^+_0)$.
Let $r<1-s$ and $f\in C_r^0(\mathbb{R}_0^+)$ with $D^sf\in C^0(\mathbb{R}^+)\cap L^1_{loc}(\mathbb{R}^+)$. Then $J^sD^sf=f$.

\bigskip

Finally, we list some properties for the Caputo derivative:
\begin{itemize}
\item $\hat{D}f=Df$

\smallskip

\item If $f(0)=0$ then $\hat{D}^{\alpha}f=D^{\alpha}f$ and 
$\hat{D}\hat{D}^\alpha f = \hat{D}^\alpha\hat{D} f =\hat{D}^{1+\alpha} f$
\end{itemize}

\section{Generalized constitutive equations in Electromagnetism}

\subsection{Curie's law}
Consider a dielectric material and apply at $t=0$ a constant dc voltage $v_0$. The current produced is
\begin{equation}
i(t)=\frac{v_0}{ht^\alpha}, \quad 0<\alpha<1,\quad t>0.
\label{Curie}
\end{equation}

The constant $h$ is related to the capacitance of the capacitor and the kind of dielectric.  Whereas the constant $\alpha$ is associated to the losses of the capacitor. The  losses decrease with $\alpha$ approaching $1$.

\bigskip

Equation \eqref{Curie} is known as \emph{Curie's law}, see Curie \cite{MJCurie}.

\bigskip

Let $v(t)=v_0H(t)$, where $H(t)$ is the Heaviside function. In terms of the Riemann-Liouville derivative, Curie's law reads
\begin{equation}
i(t)=CD^\alpha v(t), \quad C=\frac{\Gamma(1-\alpha)}{h}.
\label{DCurie}
\end{equation}

\bigskip

Based on exhaustive experimental evidence, Westerlund \& Ekstam \cite{Cour} contend that all dielectrics and insulators follow Curie's Law. We may consider this phenomenon as a particular case of transport in complex media.  Consequently, models involving fractional derivatives seem appropriate. More on this below.

\bigskip

\noindent\textbf{Remark. } As noted above, the definition of fractional derivative is not unique. Besides the fractional derivatives of Riemann-Liouville and Caputo  there are other definitions, one commonly used is that of Gr\"{u}nwald-Letinkov. A case for the Caputo derivative is that for fractional differential equations, initial conditions are straightforward. In particular, the Caputo derivative of constant functions is zero. The latter precludes the use of Caputo derivative for describing Curie's law in terms of fractional derivatives. Consequently, our choice of the Riemann-Liouville derivative is necessary.

\subsection{Generalized constitutive equations}

Let $\mathbf{f}:\mathbb{R}^3\times\mathbb{R}^+\to \mathbb{R}^3.$ Thus $\mathbf{f}(\mathbf{\cdot},t)$ is a vector field.
Let us denote by $J_t^{1-\alpha}\mathbf{f}$  the Riemann-Liouville integral of $\mathbf{f}$ applied component wise. Similarly, the Riemann-Liouville and Caputo fractional derivatives are denoted as $D_t^{\alpha}\mathbf{f}$ and $\hat{D}_t^{\alpha}\mathbf{f}$, respectively. As before when dealing with fractional time derivatives, we consider $\mathbf{f}(\mathbf{x},t)\equiv\mathbf{f}(\mathbf{x})H(t)$, for $\mathbf{f}$ independent of $t$ .

\bigskip

Let us recall Maxwell's Equations in (macroscopic) continuous media,

\begin{align}
\nabla\cdot\mathcal{D}&=\rho\label{EqMax1}\\
\nabla\times\mathcal{E}&=-\frac{\partial\mathcal{B}}{\partial t}\label{EqMax2}\\
\nabla\cdot\mathcal{B}&=0\label{EqMax3}\\
\nabla\times\mathcal{H}&=\frac{\partial \mathcal{D}}{\partial t}+\mathcal{J}\label{EqMax4}
\end{align}

\bigskip

The physical quantities are the electric field $\mathcal{E}$, the magnetic field $\mathcal{B}$, the electric flux density $\mathcal{D}$, the magnetic field intensity $\mathcal{H}$, electric current density $\mathcal{J}$, and electric charge density $\rho$.

\bigskip

To have a complete system of equations, constitutive relations are necessary for a specific medium. These are relations for $\mathcal{D}$, $\mathcal{H}$ and $\mathcal{J}$ in terms of $\mathcal{E}$, $\mathcal{B}$. For complex material media, e.g. lossy media, these relations need not be simple. They may depend on past history, may be nonlinear, etc.

\bigskip

Consider Ohm's law 
\[
\mathcal{J}=\sigma\mathcal{E},\quad \sigma\geq 0,
\]
with $\sigma$ the positive \emph{constant of conductivity} of the medium considered. This law is valid in isotropic and homogeneous media, without memory.

\bigskip

Curie's law \eqref{Curie} indicates that there is loss of energy in the process under study, and the fractional differential law \eqref{DCurie} models a medium with memory. 

\bigskip

Consequently,  for lossy media in Electromagnetism we propose the generalised Ohm's law
 
 \begin{align}
\mathcal{J}(\mathbf{x},t)&=\sigma_{\alpha}D_t^{1-\alpha}\mathcal{E}(\mathbf{x},t),\quad 0<\alpha\leq 1.
\label{OhmF2}
\end{align}
When $\alpha=1$ the classical law is recovered. 

\bigskip

\noindent\textbf{Remark. }Law \eqref{OhmF2} is in analogy to a generalized Darcy law in Fluid Mechanics. Indeed,  the \emph{continuity equation} for electric charges, derived  from equations \eqref{EqMax1} and \eqref{EqMax4} reads 
\begin{align}
\frac{\partial \rho}{\partial t}+\nabla\cdot\mathcal{J}=0.\label{Econt}
\end{align}

\bigskip

If $\rho$ is fluid density, Equation~\eqref{Econt} is the continuity equation in Fluid Mechanics. For highly heterogeneous media, anomalous behaviour has been observed, such as non darcian flow. Generalized laws have been proposed to better describe this behaviour. If $\mathbf{u}(\mathbf{x},t)$ is the superficial velocity and $P(\mathbf{x},t)$ is the pressure, the generalized Darcy law 
\begin{align}
\mathbf{u}(\mathbf{x},t)=-\frac{k_{\alpha}}{\mu}\Gamma(\alpha)D_t^{1-\alpha}\nabla P(\mathbf{x},t),\quad 0<\alpha\leq 1\label{Darcy}
\end{align} 
has proven useful. See Moreles et al.~\cite{Mor} and references within.

\section{Fractional Circuit Elements}

Here we follow the classical derivation of the RCL circuit components, while introducing generalized constitutive relations for $\mathcal{D}$ and $\mathcal{B}$ analogue to \eqref{OhmF2}.

\subsection{Fractional Resistor}

Consider  a lossy material in the form of a cylinder of length $l$ and cross section $S$ of area $A$. Assume a constant electric field in the direction of the axis of the cylinder $\mathbf{z}$. The voltage $v(t)$ between the ends of the cylinder is 
\begin{align}
v(t)&=\int_0^l\mathcal{E}\cdot \mathbf{z}ds\notag\\
&=l\mathcal{E}\cdot \mathbf{z}\label{DCourie1}
\end{align}
whereas the total current $i(t)$ is
\begin{equation}
i(t)=\int_S\mathcal{J}\cdot \mathbf{z} dS
\label{DCourie2}
\end{equation}

Assume \eqref{OhmF2} is valid. We have
\[
i(t)=A \sigma_\alpha D^{1-\alpha}_t\mathcal{E}\cdot \mathbf{z}.
\]

Hence
\[
i(t)=\frac{A \sigma_\alpha}{l}D^{1-\alpha}_t v(t)
\]

\bigskip

Let $R_{\alpha}= \frac{l}{A\sigma_\alpha}$. It follows that the fractional resistor satisfies the Curie's law

\begin{equation}
i(t)=\frac{1}{R_{\alpha}}D_t^{1-\alpha}v(t)
\label{DCourie3}
\end{equation}

Notice that if $\alpha =1$, we have the classical law
\[
v=Ri.
\]

\subsection{Fractional Capacitor}

Consider two parallel plates confining a lossy material. Both plates have charge of equal magnitude but opposite sign. If distance between the plates is small compared with their size, there is no charge in the region between the plates. Charge will reside mostly in the inner surfaces of the plates. The electric field is normal to the plates away from the edges, say in the $x$ direction, and zero in the interior of the plates.

Consequently, in the region between the plates equation \eqref{EqMax1} becomes

\[
\nabla\cdot\mathcal{D}=0.
\]

\bigskip

We propose the constitutive law
\begin{equation}
\mathcal{D}=\epsilon_{\beta}D_t^{1-\beta}\mathcal{E},\quad 0<\beta\leq 1.
\label{DEFrac}
\end{equation}

\bigskip

Since $\mathcal{E}=-\nabla\phi$, the potential $\phi(x,y,z)\equiv\phi(x)$ solves the equation

\begin{align}
D_xD_t^{1-\beta}D_x\phi=0\label{DCap2}
\end{align}
We have
\[
D_x\left(\frac{1}{\Gamma(\beta)}t^{\beta-1}D_x\phi\right)=0,
\]
and for $t>0$
\[
D^2_x\phi=0.
\]

\bigskip

Let the plates be at $x=a$ and $x=b$. As customary, suppose that the potentials are constant on each plate and are $\phi(a)=v_a$ and $\phi(b)=v_b$.
Then
\[
\phi(x)=\frac{(v_b-v_a)x+bv_a-av_b}{b-a},
\]
and the electric field
\begin{equation}
\mathcal{E}=\frac{v_{ab}}{b-a},
\label{EFCap}
\end{equation}
where $v_{ab}=v_a-v_b$.

\bigskip

Now choose $\delta$ such that $\delta<<|b-a|$ and construct the cylindrical surface $S_C$, with axis in the $x$-direction. The plane top and bottom, of area $A$, are at $x=a+\frac{\delta}{2}$ and $x=a-\frac{\delta}{2}$.

From the integral form of equation~\eqref{EqMax1} in the region confined by $S_C$ we obtain

\begin{equation}
qA=\int_{S_C}\mathcal{D}\cdot n\, ds,
\end{equation}
where $q$ is the constant surface density of charge, positive on $x=a$.

Let $v(t)=v_{ab}H(t)$ and $Q(t)$ the total charge on the plate at $x=a$, namely $Q(t)=qAH(t)$. From \eqref{DEFrac} and \eqref{EFCap} we have
\[
Q=\frac{\epsilon_{\beta}A}{b-a}D_t^{1-\beta}v
\]

\bigskip

We are led to a  generalized governing equation of a capacitor

\begin{equation}
Q(t)=C_{\beta} D_t^{1-\beta}v(t)
\label{DCap6}
\end{equation}
with
\[
C_{\beta}=\frac{\epsilon_{\beta} A}{b-a}.
\]

\subsection{Fractional Inductor}

For later reference let us recall Ampere's law and Faraday's law of induction. Namely

\begin{align}
i(t)&=\oint_C\mathcal{H}\cdot \mathbf{t} dl\label{Ampere}\\
\oint_C  \mathcal{E}\cdot \mathbf{t} dl &=-\frac{d}{dt}\int_S\mathcal{B}\cdot \mathbf{n} ds\label{Faraday}
\end{align}

The corresponding generalized constitutive relation between the magnetic field intensity $\mathcal{H}$ and the magnetic field $\mathcal{B}$ is

\begin{equation}
\mu_{\gamma}\mathcal{H}=D^{1-\gamma}_t\mathcal{B},\quad 0<\gamma\leq 1.
\label{FHB}
\end{equation}

For the inductor it is considered a toroidal frame of rectangular cross section. The inner and outer radii of the frame are $r_1$ and $r_2$ respectively, and the height is $h$.   A coil consisting of $n$ turns of wire is tightly wound on the frame.  There is a current of magnitude $i(t),t \geq 0$ in the conducting wire.

Set cylindrical coordinates such that the $z-$axis is the axis of symmetry and the frame is located between $r=r_1$ and $r=r_2$. Let $C_r$ be a circular path within the toroidal farme of radius $r$ with $r_1<r<r_2$.  As customary, we assume axis symmetry so that the magnetic field $\mathcal{B}\equiv\mathcal{B}(r,z,t)$ only depends on $r,z,$ and $t$.

By Ampere's law applied to the surface of the disk bounded by $C_r$, we have
\[
ni(t)=\int_{C_r} \mathcal{H}\cdot\mathbf{t}dl,
\]
where $\mathbf{t}$ is the unit tangent to $C_r$.

Substituting $\mathcal{B}$ from \eqref{FHB} we obtain
\[
\mu_\gamma ni(t)= D^{1-\gamma}_t\int_{C_r} \mathcal{B}\cdot\mathbf{t}dl.
\]
After application of the Riemann-LIouville integral of order $1-\gamma$ we are led to
\[
\mu_\gamma nJ^{1-\gamma}_ti(t)= \int_{C_r} \mathcal{B}\cdot\mathbf{t}dl.
\]

If $B$ is the magnitude of the magnetic field in the direction of $\mathbf{t}$, then
\[
\int_{C_r} \mathcal{B}\cdot\mathbf{t}dl=\int_0^{2\pi}B(r,z,t)dl=2\pi rB(r,z,t).
\]
So
\[
B(r,z,t)=\frac{\mu_\gamma n}{2\pi r}J^{1-\gamma}_ti(t).
\]
 Let us apply Faraday's law to one such surface $S$. Then the normal to this surface is $\mathbf{n}=\mathbf{t}$. So if $v_S(t)$ is the induced voltage

\begin{eqnarray*}
v_S(t) & = & -\frac{d}{dt}\int_S \mathcal{B}\cdot\mathbf{n}dS \\
  & = & -\frac{d}{dt}\int_0^h\int_{r_1}^{r_2} B(r,z,t)drdz\\
  & = & -\frac{\mu_\gamma n}{2\pi }\int_0^h\int_{r_1}^{r_2} \frac{1}{r}drdz\frac{d}{dt}J^{1-\gamma}_ti(t)\\
  & = & -\frac{\mu_\gamma nh}{2\pi }log(r_2/r_1)D_t^{\gamma}i(t).
 \end{eqnarray*}

\bigskip

Let $v$ be the total voltage dropped. Since there are $n$ such coils, we have
\begin{equation}
v(t)=L_{\gamma}D_t^{\gamma}i(t)
\label{Finductor}
\end{equation}
with $L_{\gamma}=\frac{\mu_\gamma n}{2\pi }log(r_2/r_1)$.

\bigskip

Equation~\eqref{Finductor} is a generalization of the governing equation of an inductor.

\section{Fractional RCL Circuit}

Table~\ref{tab:DiF} summarizes the equations obtained in the previous section. In the second column the fractional generalizations are shown, while the third column presents the limit behavior of the fractional models. As expected the classical governing equations, representing ideal components of a circuit, are obtained.

\bigskip

\begin{table}[h]
\centering
\begin{tabular}{ccc}
\toprule[1.5pt]
\textbf{Component}&\textbf{Fractional Equation}&\textbf{Classical Equation}\\
\midrule
Resistor&$i(t)=\frac{1}{R_{\alpha}}D_t^{1-\alpha}v(t)$&$i(t)=\frac{1}{R}v(t)$\\
Capacitor&$Q(t)=C_{\beta} D_t^{1-\beta}v(t)$&$Q(t)=Cv(t)$\\
Inductor&$v(t)=L_{\gamma}D_t^{\gamma}i(t)$&$v(t)=LD_t[i(t)]$\\
\bottomrule[1.5pt]
\end{tabular}
\caption{Fractional and classical governing equations of circuit components}\label{tab:DiF}
\end{table}

For $v$ regular enough we have for the fractional resistor
\[
v(t)=R_{\alpha}J_t^{1-\alpha}i(t),
\]
whereas for the fractional capacitor
\[
v(t)=\frac{1}{C_\beta}J_t^{1-\beta}Q(t).
\]

\begin{figure}[h]
\centering
\includegraphics[width=0.5\textwidth]{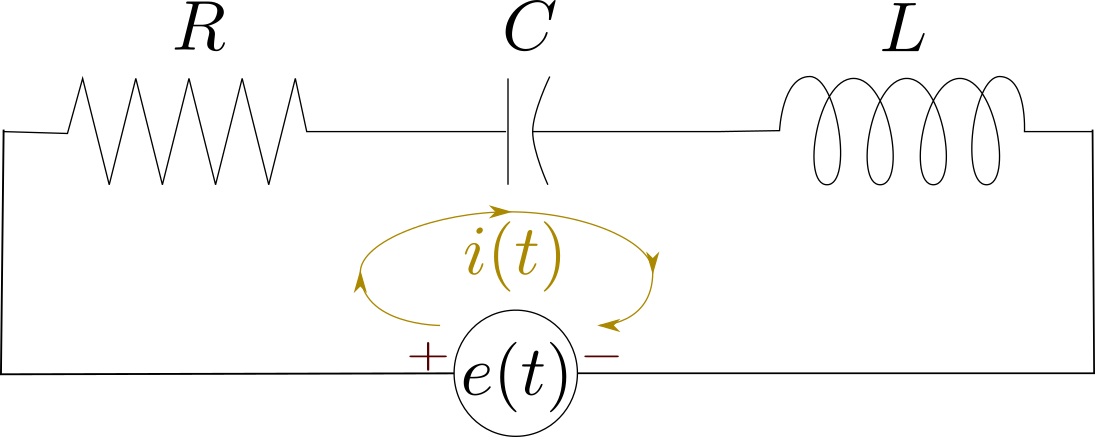}
\caption{Fractional RCL circuit}\label{fig: CirFrac}
\end{figure}

Consider a resistance, a capacitor, an inductor in series with a power source $e(t)$, as shown in Figure~\ref{fig: CirFrac}. The first three components have fractional governing equations.

By Kirchhoff's law of voltages we have:
\begin{equation*}
e(t)=R_{\alpha}J_t^{1-\alpha}i(t)+\frac{1}{C_\beta}J_t^{1-\beta}Q(t)+L_{\gamma}D_t^{\gamma}i(t),
\end{equation*}
but from the continuity equation \eqref{Econt} it is known that $i(t)=D_tQ(t)$. We are led to the Fractional RCL circuit equation
\begin{equation}
L_{\gamma}D_t^{\gamma}D_tQ(t)+R_{\alpha}J_t^{1-\alpha}D_tQ(t)+\frac{1}{C_\beta}J_t^{1-\beta}Q(t)=e(t).
\label{CirFrac1}
\end{equation}

\bigskip

In terms of Caputo derivatives
\begin{equation}
L_{\gamma}D_t\hat{D}_t^{\gamma}Q(t)+R_{\alpha}\hat{D}_t^{\alpha}Q(t)+\frac{1}{C_\beta}J_t^{1-\beta}Q(t)=e(t).
\label{CirFrac1C}
\end{equation}

\bigskip

This is a fractional integro-differential equation of order $1+\gamma$ in $t$, with derivatives in the sense of Caputo. Thus, $Q(0)$ and $D_tQ(0)$ are the natural initial conditions.

\bigskip

If $Q(t)$ is regular enough, then $J_t^{1-\beta}Q(0)=0$ and the Riemann-Liouville and Caputo derivatives coincide. Applying the Caputo derivative of order $1-\beta$ in \eqref{CirFrac1C} we obtain

\begin{equation}
L_{\gamma}\hat{D}_t^{1-\beta}D_t\hat{D}_t^{\gamma}Q(t)+R_{\alpha}\hat{D}_t^{1-\beta}\hat{D}_t^{\alpha}Q(t)+\frac{1}{C_\beta}Q(t)=\hat{D}_t^{1-\beta}e(t).
\label{CirFrac1C2}
\end{equation}

\bigskip

This is the form commonly found in applications. See for instance Magin \cite{Magin}.

\end{document}